\documentclass[prd,twocolumn,showpacs,preprintnumbers,amsmath,amssymb,superscriptaddress]{revtex4-1}

\usepackage{graphicx,subfigure}
\usepackage{dcolumn}
\usepackage{bm}

\def\nuebar{\bar{\nu}_e}
\graphicspath{{./figures/}}

\begin{document}

\preprint{Physical Review C}

\title{SNIF: A Futuristic Neutrino Probe for Undeclared Nuclear Fission Reactors}

\author{Thierry Lasserre}
\email{Corresponding author: thierry.lasserre@cea.fr}
\affiliation{Commissariat \`a l'Energie Atomique et aux Energies Alternatives,\\
Centre de Saclay, IRFU/SPP, 91191 Gif-sur-Yvette, France}

\author{Maximilien Fechner}
\affiliation{Commissariat \`a l'Energie Atomique et aux Energies Alternatives,\\
Centre de Saclay, IRFU/SPP, 91191 Gif-sur-Yvette, France}

\author{Guillaume Mention}
\affiliation{Commissariat \`a l'Energie Atomique et aux Energies Alternatives,\\
Centre de Saclay, IRFU/SPP, 91191 Gif-sur-Yvette, France}

\author{Romain Reboulleau}
\affiliation{Ecole Polytechnique, Palaiseau, France}
\affiliation{Commissariat \`a l'Energie Atomique et aux Energies Alternatives,\\
Centre de Saclay, IRFU/SPP, 91191 Gif-sur-Yvette, France}

\author{Michel Cribier}
\affiliation{Commissariat \`a l'Energie Atomique et aux Energies Alternatives,\\
Centre de Saclay, IRFU/SPP, 91191 Gif-sur-Yvette, France}

\author{Alain Letourneau}
\affiliation{Commissariat \`a l'Energie Atomique et aux Energies Alternatives,\\
Centre de Saclay, IRFU/SPhN, 91191 Gif-sur-Yvette, France}

\author{David Lhuillier}
\affiliation{Commissariat \`a l'Energie Atomique et aux Energies Alternatives,\\
Centre de Saclay, IRFU/SPhN, 91191 Gif-sur-Yvette, France}

\date{\today}

\begin{abstract}
Today reactor neutrino experiments are at the cutting edge of
fundamental research in particle physics. Understanding the
neutrino is far from complete, but thanks to the impressive progress
in this field over the last 15 years, a few research groups are
seriously considering that neutrinos could be useful for society.  The
International Atomic Energy Agency (IAEA) works with its Member States
to promote safe, secure and peaceful nuclear technologies. In a
context of international tension and nuclear renaissance, neutrino 
detectors could help IAEA to enforce the Treaty on the
Non-Proliferation of Nuclear Weapons (NPT). 
In this article we discuss a futuristic neutrino application to detect and localize an
undeclared nuclear reactor from across borders.  The
SNIF\footnote{Secret Neutrino Interactions Finder} concept proposes
to use a few hundred thousand tons neutrino detectors to unveil
clandestine fission reactors. Beyond previous
studies we provide estimates of all known background sources as a function
of the detector's longitude, latitude and depth, and we discuss
how they impact the detectability.
\end{abstract}

\maketitle

\section{Neutrinos and Proliferation}
In a context of increasing carbon-free emission energy needs, civilian
nuclear energy will probably expand all over the world. Globalization, as well
as the goal of energy independence, led to an increase of the list of
countries aiming to acquire technological know-how in the field of nuclear
energy. As a consequence of the spread of practical knowledge, the
possibility of diverting a nuclear facility towards non-civilian use
could increase in the next 50 years. The main duty of the United Nations' International
Atomic Energy Agency (IAEA) is to work to make sure that nations use
nuclear energy only for peaceful purposes~\cite{aieawww}. Apart from
political difficulties regarding safeguards, the efficiency of
IAEA controls may be limited by monitoring techniques in the
future due to the fast growth of nuclear facilities around the
world. Since 2003, the Department of Safeguards of the International
Atomic Energy Agency has been evaluating the potential applicability of
neutrino detection technologies for safeguard purposes.

In 2008, a transverse working group of reactor neutrino experts from
the Member States together with the IAEA Division of Technical
Support (SGTS) firmly established that neutrino detectors have
unique abilities to non intrusively monitor a nuclear reactor's
operational status, power and fissile content in real-time, from
outside the reactor containment. This led to the definition of three safeguards
scenarios of interest. The first two, the confirmation of the absence
of unrecorded production of fissile material in declared reactors
and the estimation of the total burn-up of a reactor core,
are related to the so-called near-field applications with detectors
located a few tens of meters from the core. The third scenario
concerns clandestine or undeclared nuclear reactor detection with
core-detector distances ranging from tens to hundreds of kilometers,
also known as far-field applications.  

As far as near-field monitoring is concerned, a few detectors
specifically built for safeguards have shown robust, long term
measurements of these metrics in actual installations at operating
power reactors \cite{songs, nucifer}. Several experimental
programs~\cite{aiea2008report} are currently being carried out in
Brazil, France, Italy, Japan, Russia and the United States, guided by IAEA inputs on
their needs at specific reactors.  Over a longer time scale, it has
been recognized that neutrino detectors could have a considerable
value in bulk process and safeguards by design approaches for new and
next generation reactors. Concerning far-field applications, an 
undeclared reactor operating secretly  would stand out in the received
$\nuebar$. Such a detection possibility has already been discussed 
in~\cite{learned, guillian}, using a network of gigantic water
Cherenkov detectors (of 1 million tons each) being deployed below four
kilometers of water in deep oceans. 

The purpose of this article is to address the possibility of detecting 
undeclared nuclear reactors across borders with very large neutrino 
detectors, outlining basic principles and figures regarding the
deployment of large neutrino detectors as a safeguard tool. 

Our study revisits the detectability of clandestine nuclear reactors
with a comprehensive treatement of known neutrino sources and
backgrounds. In this article we consider as a baseline case a
10$^{34}$ free protons liquid scintillator detector fitting inside an
oil supertanker. 
The strategy presented in this article is to deploy one or more of
these detectors as close
as possible to a suspicious area, typically between 100~and 500~km. 
The detectors are then temporarily sunk underwater for data taking
until a significant amount of events are detected. Our baseline
operating time is 6 months.

We will first introduce the reactor neutrino field in section \ref{sec:nusources}.  
In section \ref{detection} we review the neutrino technology 
and we outline the detector design in section~\ref{detlayout}.
We then compute the known neutrino sources (section~\ref{neutrinobkg}) and backgrounds
(section~\ref{nonnubkg}).
We then establish the rogue activity detection criteria in section~\ref{nuclearreactordetection}. 
In section~\ref{nonnubkg} we study non-neutrino backgrounds
 in order to derive and possibly relax the minimum operation depth 
with respect to previous studies~\cite{guillian}. 
We also provide a method for determining the location of an undeclared reactor
(section~\ref{reactorloc}). We conclude by addressing the feasibility of the 
 concept within the next thirty years.

%
\section{Reactor neutrinos}
\label{sec:nusources}
\subsection{Neutrino production}
Fission reactors are prodigious producers of neutrinos, emitting about
$10^{21}$ $\nuebar$ $\mathrm{s}^{-1}$ $\mathrm{GW}_{\rm th}^{-1}$. 
In this article we will only consider the electron antineutrinos
emitted by $\beta$-decays, referred to simply as neutrinos in the rest of
the text. 
In actual light water reactors, the Uranium fuel is enriched to a few percent in~$^{235}$U.
The fission of $^{235,238}$U and $^{239,241}$Pu isotopes
produces neutron rich nuclei which must shed neutrons
to approach the valley of stability. The beta decays of these fission products
produce approximately six electron antineutrinos per fission. 
Measurements for $^{235}$U and $^{239,241}$Pu and
theoretical calculations for $^{238}$U are used to evaluate the
$\nuebar$ spectrum \cite{Schreckenbach:1985ep,Hahn:1989zr,lm2010}. Since
$^{238}$U only contributes to about 11\% of the neutrino signal, and
further since the error associated is less than 10\%, $^{238}$U
contributes less than 1\% to the overall uncertainty in
the~$\nuebar$~flux.  The overall normalization is known to about
$1.4\%$ \cite{Declais} and its shape to about $2\%$ \cite{Bugey}. As a
nuclear reactor operates, the proportions of the fissile elements evolve with
time. During a typical fuel cycle the Pu concentrations increase, so
the total neutrino flux decreases with time. 
As an approximation we use a typical averaged fuel composition during a
reactor cycle corresponding to the following fission fractions:
$^{235}$U (55.6\%), $^{239}$Pu (32.6\%), $^{238}$U (7.1\%) and $^{241}$Pu (4.7\%).
\subsection{Neutrino oscillations}
There is now compelling evidence for flavor conversion, also
known as oscillations, of atmospheric, solar, reactor and accelerator
neutrinos~\cite{SK98,osc2,osc3,sno,KamLAND1,K2K,Minos}. 
Thus reactor neutrino experiments measure 
a rate weighted by the survival
probability~$P(\nuebar\rightarrow\nuebar)$ of
the~$\nuebar$ emitted by nuclear power stations at a
distance~(L), resulting in a deviation from the $1/L^2$ dependence that
would otherwise be expected.

Reactor neutrino oscillations depend on the atmospheric
$\Delta m_{31}^{2}$ and the solar $\Delta m_{21}^{2}$
mass splittings between the three neutrino mass eigenstates, as well
as the three mixing angles $\theta_{12}$, $\theta_{23}$, and the
small, still undetermined $\theta_{13}$~\cite{PDG}. In this article, we
consider minimum baselines of about 100~km. Because $\Delta
m_{sol}^{2}\ll\Delta m_{atm}^{2}$ and because of the smallness of
$\theta_{13}$, the oscillation probability can be approximated
by: \[1-P(\nuebar\rightarrow\nuebar)=\sin^{2}(2\theta_{12})\,\sin^{2}\left(1.27\frac{\Delta
  m_{21}^{2}[{\rm eV}^{2}]L[{\rm m}]}{E_{\nuebar}[{\rm MeV}]}\right)\:\] 
For energies above 1.8~MeV, the survival probability is close to 1
 for distances of 0 to several tens of kilometers. The
survival probability then oscillates around an asymptotic value of 0.57 as the
distance ranges from about 50~km to 300~km. At further distances, much larger
than the 'solar-driven' oscillation length, the probability is
practically very close to 0.57. In this work we treat neutrino oscillation with the
state-of-the-art three neutrino oscillations formula~\cite{PDG}, but we
set $\theta_{13}$ to 0 since its small value will not impact the oscillation
probability for the purpose of this study.

Because of the combination of MeV range energies
and baselines less than $10^3$~km the modification of
the oscillation probability induced by the coherent forward scattering
 from matter electrons is less than a few
 percent. In this work the effect is small enough to be neglected.
\subsection{Neutrino rate}
\label{sec:nurates}
\begin{figure}[!h]
\begin{center}
        \includegraphics[scale=0.45]{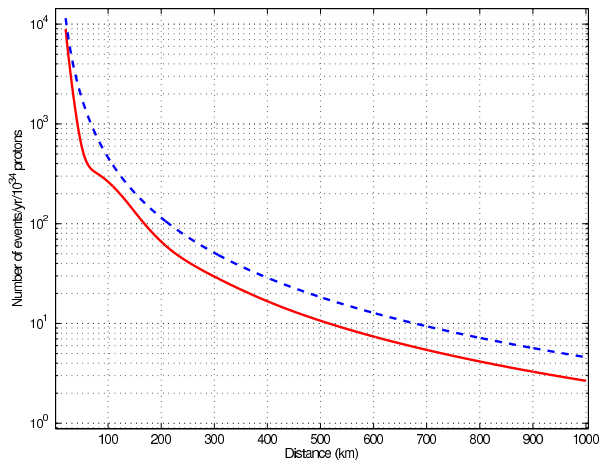}
	\caption{\label{num} Number of events detected in 1 year in a 
detector with $10^{34}$ protons, as a function
of the distance from a single 100~MW reactor. 
The dashed line shows the variation in the absence 
of neutrino oscillations, while the full line shows the 
actual observation taking into account oscillations.
 From 50 km to 100~km a caracteristic 'wiggle' can be seen. 
Beyond 250 km, oscillations simply cause a reduction 
in the flux, but the $1/L^2$ dependence is restored.}
\end{center}
\end{figure}

The mean energy released $\left<E_f\right>$  per fission is 
around 205 MeV~\cite{kopeikin}. The energy-weighted cross section 
to detect neutrinos amounts to 
$\left<\sigma_f\right>= 5.8 \times 10^{-43}$~cm$^2$ per fission. The
reactor thermal power ($P_{\rm th}$) is related to the number of fissions
per second ($N_f$) by $N_f = 6.24 \times 10^{18} \; {\rm s}^{-1} \,
P_{\rm th}[{\rm MW}] / \left<E_f\right>[{\rm MeV}]$. The neutrino
interaction rate ($R_L$) at a distance $L$ from the source, assuming 
no neutrino oscillations, is then
\mbox{$R_L=N_f \left<\sigma_f\right> n_p / (4\pi L^2)$},
where $n_p$ is the number of hydrogen atoms, or free protons, of
the target.
The expected neutrino rate from clandestine reactors at a few 
hundred kilometers is quite small, due to the weak neutrino 
interaction cross section and the $1/L^2$ dependence.
 Neglecting neutrino oscillations, a 100~MW$_{\rm th}$ 
reactor would induce 450~events per year in a $10^{34}$~proton detector
at a distance of 100~km. In this particular case, neutrino oscillations induce a
deviation from the $1/L^2$ dependence and only 250 neutrino events
 could effectively be detected as shown on Figure~\ref{num}. 
Therefore the target mass must at least be of the order of a
hundred thousand tons. For fundamental research, neutrino detectors 
as large as 50,000~tons have been built \cite{SK98}. 
A few projects of larger liquid scintillators detectors, like 
\cite{lena, hanohano}, are currently being discussed. 
\subsection{Clandestine reactors}
We define clandestine or rogue reactors as nuclear reactors not
declared by a country to the international community
(IAEA). Like regular research or power reactors they would be copious
sources of neutrinos. 
We assume such reactors to have the same features as regular
reactors, though their nuclear thermal power is unknown. In this
article we will consider that clandestine reactors may have powers
between 10~MW$_{\rm th}$ to 2~GW$_{\rm th}$. 
We consider the fuel composition of clandestine reactors to be similar
 to the composition of commercial reactors, on average.
 This assumption has no strong impact
on the detectability of clandestine activity since the neutrino
rate depends mainly on the thermal power.
\section{Neutrino detection}
\label{detection}
Electron antineutrinos can be detected \textit{via} inverse beta
decay on free protons: $\nuebar + p \to e^+ + n$ for $E_{\nuebar} > E_{\rm thr}
\sim 1.8$~MeV, and their energy is derived from the measured positron
kinetic energy as $E_{\nuebar} \simeq E_{e^+} + E_{\rm thr}$. We call
\textit{visible energy} the energy deposited in the detector, corresponding to
$E_{\rm vis} = E_{e^+} + 2 m_e$. The inverse
beta decay cross section has been precisely computed in
\cite{VogelBeacom}.
A neutrino event is thus characterized by a prompt positron event which
deposits a visible energy between 1 and 8~MeV, followed by a delayed
gamma event arising from neutron capture within $\tau \sim 10-200\;\mu$s.
The minimal energy of 1~MeV of the prompt event is caused by the
positron's annihilation in the active volume. Prompt and delayed event
are spatially correlated, within $< 1 \; {\rm m}^3$. They both have a 
$\beta / \gamma$-type pulse shape, distinguishable in liquid
scintillators. This signature allows to discriminate efficiently
against backgrounds.

Water-Cherenkov and liquid scintillator detectors allow real-time
spectroscopy of electron antineutrinos.

Liquid scintillators have commonly been used for the past 60 years to
detect neutrinos using the inverse $\beta$-decay reaction. The
hydrogen atoms serve as targets to neutrinos, producing ionizing
particles. The liquid scintillator emits light in the UV-range when
crossed by these charged particles. These liquids can be flammable or
dangerous to the environment and thus require special care when large
amounts are handled. The inverse
beta-decay reaction does not allow to recover the direction of the
incoming neutrino, apart from a slight backward shift of the positron
angular distribution. The scintillation light, isotropically emitted, allows to
find the position of the neutrino interaction within a few tens of
centimeters. Large volume detectors may yield a few hundred
photoelectrons per deposited MeV, corresponding to an energy
resolution of a few percent even for the less energetic neutrinos.
This detection technology based on liquid scintillator has
the capability to measure the full $\nuebar$ spectrum since the
instrumental threshold may be lowered to 1~MeV or less, depending on
backgrounds.

High purity water is also used as a detection medium for charged
particles traveling through at super-luminous speed~\cite{SK98},
inducing so-called Cherenkov radiation. Water has a few advantages:
it is straightforward to handle, non flammable, non toxic, available in
large volumes at relatively low cost, and easily purified by common
techniques to improve its transparency. For charged particles above
an energy threshold (0.78~MeV for electrons) only 200~UV photons/cm 
are emitted along the track, i.e. roughly 30~times less light than liquid
scintillators. Similarly this detection technique cannot be used to
determine the direction of the incoming neutrino.

For both technologies, doping the liquid with Gadolinium at the level
of 1--5~g/l can greatly improve the sensitivity to electron
antineutrinos from reactors. The large cross section for neutron
capture on $^{157}$Gd ($2.59\,10^5$~barn) and $^{155}$Gd ($6.1
\, 10^4$~barn) enhances the sensitivity to the delayed neutron
signal. The positron, emitting scintillation photons or radiating
Cherenkov photons, is immediately detected with or without
Gadolinium. However the neutron, quickly thermalized in the hydrogen-rich
media, is captured on Gd with a probability of more than
80\%. Upon capturing a neutron, a Gadolinium nucleus relaxes to its
ground state by emitting a cascade of gamma rays having a total energy
of about 8~MeV, thus enhancing the detection efficiency. This is
especially true in water where the neutron capture signal on hydrogen,
at 2.2~MeV, is barely detectable above backgrounds.  Furthermore the
time delay between the positron and neutron events is significantly
decreased, leading to a reduction of accidental backgrounds. Unlike
Gadolinium-doped water, stable Gadolinium-doped liquid scintillators
are difficult to obtain, but we assume that current technologies
being developed for the next generation of experiments~\cite{Cras2005}
will be routinely available thirty years from now.

In both cases visible-UV light is collected by photomultiplier tubes
covering the walls of the detector vessel. 
The total charge and photon arrival times allow to reconstruct the
incident neutrino's energy and interaction time. 
\section{Detector design}
\label{detlayout}
In this section we briefly describe our baseline
detector design and address the technical challenges
in realizing a SNIF detector module. 
Our design concept is similar to those developed for
large neutrino detectors proposed for fundamental research, such as
Titand~\cite{titand}, Hano-Hano~\cite{hanohano} and LENA~\cite{lena}.

We consider a detector module containing $10^{34}$ free protons
(fiducial volume). This corresponds to 138,000~tons of
linearalkylbenzene ($\text{LAB}, {\rm C}_{18}{\rm H}_{30}$) based liquid scintillator, 
contained in a volume of 160,000~m$^{3}$ (density of 0.86). 
This could be hosted in a cylindrical tank of 23~m radius and 96.5~m length.
Optical coverage of the detector walls of 20~percent
would require 17,000~photomultipliers of the Superkamiokande type (20~inches).
Depending on the total muon rate in the module, and thus on the operating
depth, the fiducial volume could be optically segmented. 
This huge detector module has to be designed to be transportable and deployable 
in the deep ocean. We propose to host the detector in a supertanker to
be transported on the detection site. 
By comparison, a modern supertanker can have a capacity of over 
400,000~deadweight tons. It would then be immersed at a depth of
a several hundred to several thousand meters. 

In the rest of the article we will focus our discussion on liquid scintillator technology. 
 The components of the detector are embedded in each other, the liquid
scintillator being in the central volume. Linearalkylbenzene is currently used in
neutrino experiments (for instance SNO+~\cite{sno+}, Double
Chooz~\cite{dc}, Daya Bay~\cite{dayabay}) because of its good optical
 transparency ($>$ 20~m), 
its high light yield, its low amount of radioactive impurities, and
its high flash point (140 degrees Celsius) which makes safe
handling easier. Moreover experimental studies of temperature
and pressure dependence validated its performance in
deep underwater environment~\cite{hanohano}. 

Scintillator timing properties as well as optical transparency are improved by adding a
combination of solutes, typically PPO (a few g/l) and bisMSB (a few
mg/l). In addition, the neutron capture capability of the scintillator
would be greatly enhanced with the dissolution of a Gadolinium complex
(a few g/l typically). This should be considered as a serious option if long term stability
 issues are solved. It would require dedicated surface treatement
 of the inner walls, such as Teflon coating.

Such a liquid scintillator medium would lead to 80\% neutrino detection efficiency.
Setting an analysis threshold at $E_{\rm vis} > 2.6$~MeV would greatly
reduce the material radiopurity constraints, simplifying
industrial production of the modules (see Section \ref{accbkg}). We
thus consider this energy cut in our baseline scenario, and account for
the resulting loss of efficiency. 

In order to further reduce backgrounds the inner
stainless steel tank could be enclosed in another steel vessel
providing a protective layer of ultra pure water against external
radioactivity. This 1~meter thick region could be equiped with
phototubes detecting Cherenkov light from cosmic muons,
allowing to further suppress the muon-induced background
(see Section \ref{fastn}).
This would imply the installation of 4,000~additional
phototubes on the outer tank walls. 

The geometry of both tanks should be curved to accommodate
the deep-sea hydrostatic pressure. The detector should finally be
zero-buoyant to be sunk into the deep ocean for its operation and to
be brought back to the surface for maintenance or redeployment
elsewhere. 
\begin{figure*}
\begin{center}
	\includegraphics[scale=0.45]{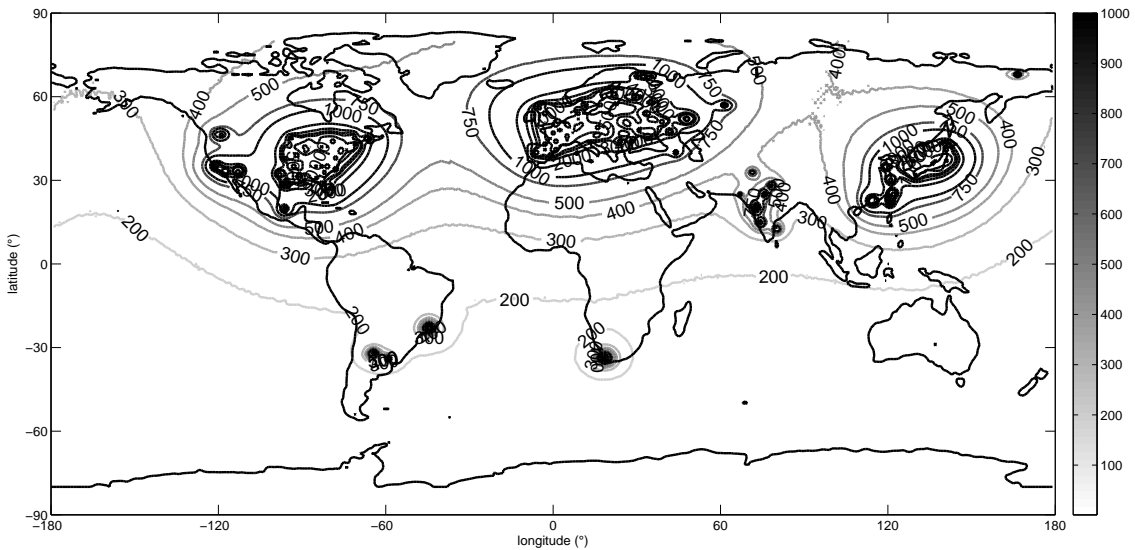}
	\caption{\label{flux} Maps illustrating the number of neutrino
          events that would be detected in a $10^{34}$ free protons
          detector
 ($E_{\rm vis} > 2.6$~MeV, 4,000~m operating depth) 
after half a year of data taking. 201 nuclear power stations 
have been included, accounting for a 78\% global load factor on averaged. 
This map includes all non-neutrino backgrounds which are negligible
 at this depth in the northern hemisphere (see Section~\ref{nonnubkg}).}
\end{center}
\end{figure*}
\section{Known neutrino sources}
\label{neutrinobkg}
In this section we review the two largest sources of known neutrinos.
These sources are an irreducible  background for the search of undeclared nuclear cores: 
power/research reactor neutrinos and geological neutrinos. 
As a first step the known sources neutrino signal should be subtracted from the observed
rate to unveil undeclared nuclear fission activity. 
\subsection{World nuclear power stations}
Clandestine reactor and commercial reactor neutrinos are totally
indistinguishable. Neutrinos from commercial plants are thus an
irreducible source of background that could overwhelm the clandestine
signal. However this background is predictable since the IAEA can access 
all power plants' geographical coordinates, thermal power, and operating
status at any time. In our simulation,
we included 201~nuclear power stations, most of them having multiple
units. They amount to 1,134~GW$_{\rm th}$ total thermal power. The reactor's
latitude and longitude were checked to a precision of one hundredth of a degree, using
satellite views \textit{via} Google Earth\textregistered. 
The mean load factor of each reactor for the 1998-2008 period has been 
included when possible. We consider that the
global power is stable, though reactors will be turning on and
off for refueling and maintenance. We assume that the day-by-day
thermal power would be knowable by the monitoring authorities, with a
3\% uncertainty. This is consistent with what present day experiments,
KamLAND and Borexino, are able to achieve
\cite{kamlandOsc2008,borexinogeonu}.
We also assume an averaged isotopic composition for all
cores known within 3\% uncertainty. 

The number of events detected in 6 months in a SNIF detector module 
as a function of geographical position on Earth 
is shown on Figure~\ref{flux}. Most of the commercial nuclear power
stations are located in the northern hemisphere, mainly belonging to
developed countries, especially the United States of America, Europe,
and Japan. These three zones gather more than 85\% of the
total world nuclear power budget.  
Only four power stations are present in the southern hemisphere,
in Argentina, Brazil, and South Africa, amounting to only 14
GW$_{\rm th}$. 
This asymmetry plays an important role in the sensitivity of the 
neutrino method. Obviously it is harder to detect clandestine activity 
in areas where commercial and/or research reactor activity is high. 

In this work we consider only power reactors
with thermal power greater than 100~MW$_{\rm th}$, and we do not account for
research reactors. This is justified because the total power produced
by commercial nuclear reactors is far greater than that of research reactors.
Though this assumption is correct for most locations around the world,
it may be locally inexact in some areas with no power stations.
\subsection{Geoneutrinos}
\label{geonu}
Geoneutrinos are natural electron antineutrinos arising from the
decay of radioactive isotopes of Uranium, Thorium, and Potassium in
the crust and mantle of the Earth. The spectrum of neutrinos from the
decay chains of Uranium and Thorium extends above the energy threshold
for inverse neutron decay (1.8 MeV) to the maximum geologic
neutrino energy (3.27 MeV), corresponding to a 2.5 MeV visible
energy deposition in a liquid scintillator detector. Potassium neutrinos are below
threshold for this reaction. For the current study we used a $2^\circ\times 2^\circ$ 
map providing the Uranium and Thorium geoneutrino fluxes based on the
Earth reference Model in~\cite{mantovanibse}. Geoneutrino fluxes
are computed following the prescription described in~\cite{geonurate},
at the detector's longitude and latitude coordinates, taking neutrino
oscillations into account. The geoneutrino background rate ranges from a few hundred
interactions per 10$^{34}$~H.year in the middle of the oceans (thin
oceanic crust), to a few thousand interactions per 10$^{34}$~H.year in
the middle of the continents (thick continental crust). A possibility
to discard this background completely is to set an analysis
threshold above 2.5 MeV. 
Another possible source of background a hypothetical natural nuclear reactor of
a few TW in the core
of the Earth \cite{georeactor}. In our study we neglect its potential influence.
\section{Non-neutrino backgrounds}
\label{nonnubkg}
Irreducible commercial reactor neutrinos are not the only source of
backgrounds. Non-neutrino backgrounds could prevent any neutrino
detection if not handled carefully since the expected low signal
(clandestine minus known activity) must not be drowned in a high rate 
of background events. 
This implies extensive passive shielding to protect the fiducial
volume from natural radioactivity, as well as active shielding to
veto cosmic rays. In addition a thickness of hundreds to thousands
meters of water is mandatory to achieve sufficient supression of
atmospheric muons, neutrons and cosmogenic radioisotopes.
 In this section we review the available detection technologies. We then
 provide a model for the three main kinds of non-neutrino backgrounds: 
accidental coincidences, fast neutrons and the long-lived muon 
induced isotopes $^9$Li/$^8$He, as a function of the operating depth.
\subsection{Accidental backgrounds}
\label{accbkg}
When detecting neutrinos, naturally occurring radioactivity (U,
Th, K) of the component of the detector may create fake signal 
- so-called accidental background -
defined as a coincidence of a prompt energy deposition between 1 and
10~MeV, followed by a delayed neutron-like event, occurring after a
delay $\tau_d$ of a few hundredths of a millisecond, in close
proximity to the prompt energy deposit (within a volume $V_d \sim
1\,\mathrm{m}^3$).  With these notations, the accidental background rate $r_{acc}$
is given by $r_{acc} \sim r_p \; r_d \; \tau_d \; V_d \; V_{det} $,
where $r_p$ and $r_d$ are the specific background rates (in units of
s$^{-1}$m$^{-3}$) for the prompt and the delayed signal,
respectively. $V_{det}$ is the total detection volume.

The potentially most dangerous of these backgrounds are those caused by
radioactive impurities within the active detection liquid. The use of
standard techniques like distillation, water extraction, nitrogen
purging, and column chromatography allows to achieve sufficiently low 
concentrations in radio--impurities~\cite{kamland0307030,borexino08}. 
The detection liquid is contained in a vessel and photomultipliers
catch the light emitted in the neutrino interaction. Those
materials and equipments also contain radioactive impurities whose
decay products may release their energy within the liquids. 
The selection of high purity materials entering the detector (mechanical structures,
photomultiplier tubes) as well as passive shielding provide an
efficient tool against this type of background. Surface/wall-induced
events could be rejected through spatial reconstruction cuts, 
with a loss of fiducial volume however. Taking accidental backgrounds into account 
leads to the detector design presented in Section~\ref{detlayout}. 
\begin{figure}[!h]
\begin{center}
	\includegraphics[scale=0.36]{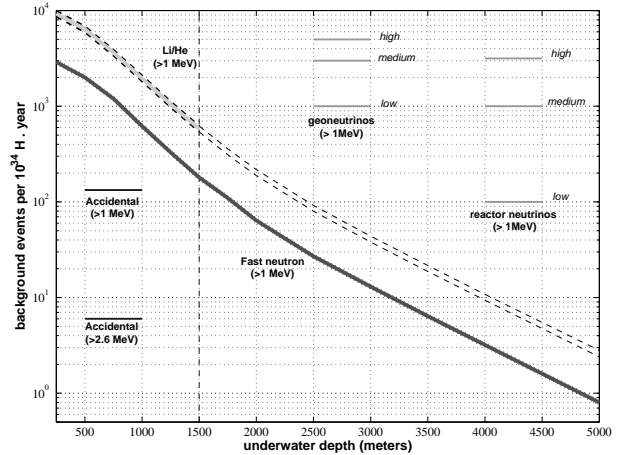}
	\caption{\label{snifbkg} Background rate as a function of the
          detector depth in meters of water equivalent for an exposure
          of $10^{34}$ H.year. Fast neutron and
          cosmogenic backgrounds are estimated with the models presented in
          sections~\ref{fastn} and~\ref{bkgcosmo}. For comparison
          we indicate our estimates of accidental backgrounds for two
          different thresholds, 1 and 2.6 MeV (visible energy). We also give three
          possible estimates for the geoneutrino rates, for
          continental crust areas (high), coastal areas (medium), and
          oceanic crust areas (low)~\cite{hanohano}. Known
          neutrino sources under three hypotheses are also
          displayed for comparison (see section \ref{neutrinobkg}).}
\end{center}
\end{figure}
In Table~\ref{tabsnifsbkg} we present our background rescaling
results.  We use the $r_p$ and $r_d$ values measured in Borexino and
KamLAND, thus the accidental background rate in SNIF scales with
$V_{det}$ only. We rescale the estimated rates for a $10^{34}$~H.y (proton.year) target immersed in deep water (SNIF baseline design). 
Choosing the Borexino extrapolation providing the lowest estimate, we
note the accidental background dominates the cosmogenics backgrounds
at depths below 2,000~m, as displayed on Figure~\ref{snifbkg}. This choice is
justified assuming a detector as radiopure as Borexino could be achievable at the SNIF
scale within the next 30~years. Considering no scaling of the delayed
signal background may be too simplistic, however. In order to get a
more robust result in our current study we get rid of the accidental
background contribution by setting an analysis threshold $E_{\rm vis} >
2.6$ MeV.  This implies a loss of signal statistics of 28\%, but it
also relaxes the radiopurity constraints by orders of magnitude,
simplifying the project feasibility.
\begin{table*}
\fontsize{8}{10}\selectfont
\begin{center}
\scalebox{1}{
\begin{tabular}{l|cc|cc}
\hline
\textbf{\hspace{1cm}Experiment} & \textbf{KamLAND} & \textbf{Borexino} & \multicolumn{2}{c}{\textbf{SNIF extrapolated from}} \\
\textbf{Label}                         &                   &          &     \textbf{KamLAND} & \&  \textbf{Borexino}  \\
\hline
Flat eq. depth & 2,050~m & 3,050~m &  \multicolumn{2}{c}{2,500~m} \\
\hline
Scintillator & \text{${\rm C}_{11.4}{\rm H}_{21.6}$}  &\text{${\rm C}_9{\rm H}_{12}$}&  \multicolumn{2}{c}{\text{${\rm C}_{16}{\rm H}_{30}$}
} \\
H/m$^3$ & $6.60\;10^{28}$& $5.30\;10^{28}$&  \multicolumn{2}{c}{$6.24\;10^{28}$} \\
C/m$^3$ &$3.35\;10^{28}$& $3.97\;10^{28}$ &  \multicolumn{2}{c}{$3.79\;10^{28}$} \\
density & 0.78 & 0.88 &  \multicolumn{2}{c}{0.86} \\
\hline
Mass (tons) & 912& 278 &  \multicolumn{2}{c}{138,000} \\
Volume (m$^3$) & 1170& 316 &  \multicolumn{2}{c}{160,000} \\
Radius (m) & 6.5& 4.25 &  \multicolumn{2}{c}{23} \\
Cyl. Length (m) &  --- & --- & \multicolumn{2}{c}{96.5}  \\
$\mu-$Section (cm$^2$) &  $1.3\;10^{6}$& $0.57\;10^{6}$ &  \multicolumn{2}{c}{$4.4\;10^{7}$} \\
\hline
$\mu-$Flux (cm$^{-2}$s$^{-1}$) & $1.6\;10^{-7}$& $0.3\;10^{-7}$ &  \multicolumn{2}{c}{$0.7\;10^{-7}$} \\
$\mu-$Energy (MeV) &219& 276 &  \multicolumn{2}{c}{247} \\
$\mu-$Rate (s$^{-1}$) & $2.13\;10^{-1}$ & $1.6\;10^{-2}$ &  \multicolumn{2}{c}{3.0} \\
$\mu-$DT (200~$\mu$s) & $4\;10^{-5}$ & $0.3\;10^{-5}$ &  \multicolumn{2}{c}{$60\;10^{-5}$} \\
Co-DT (200~ms) & $4\;10^{-2}$& $0.3\;10^{-2}$ &  \multicolumn{2}{c}{$60\;10^{-2}$} \\
\hline
Exposure (H.y) & $2.44\;10^{32}$ & $6.02\;10^{30}$ & \multicolumn{2}{c}{$10^{34}$} \\
\hline
Threshold (MeV) & 0.9 & 1 &0.9 & 1 \\
\hline
Accidental Rate & 80.5$\pm$0.1 & 0.08$\pm$0.001 & 3,300$\pm$4 & 133$\pm$2	\\
Li/He Rate &7.0$\pm$1 &0.03$\pm$0.02&85.9$\pm$12.2&108$\pm$71.9\\
Fast n Rate & 9$\pm$9& 0.025$\pm$0.025  &171$\pm$17& 93.4$\pm$9\\
Geo-$\nu$ & 69.7 & 2.5$\pm$0.2 & 2,860 & 4,150$\pm$332\\
\hline
Reactor-$\nu$ & 1,609$\pm$51 &5.7$\pm$0.3 & 65,900$\pm$2,080 & 9,460$\pm$498\\
\hline
\end{tabular}
}
\caption{Breakdown of the background estimates for SNIF. We consider
  two different background measurements from the KamLAND and Borexino
  neutrino searches~\cite{kamlandOsc2008,borexinogeonu};
  Cosmogenics from KamLAND are rescaled
  from~\cite{kamlandCosmo2010}. $\mu-$DT and Co-DT are the estimates
  of both muon and cosmogenics induced dead time (DT). The flat
  equivalent depths are taken from~\cite{meihime}. Extrapolation to SNIF
  corresponds to a 10$^{34}$~proton detector operating for 1~year at a depth of
  2,500~m, according to the prescription given in~\cite{hagner2000,meihime}. 
  In this table, SNIF is taken to be located at the KamLAND/Borexino sites 
  to calculate Geo- and Reactor-neutrino backgrounds.
  Geo-neutrino rates are measured, in agreement
  with the reference Earth model~\cite{bse}.  }
\label{tabsnifsbkg}
\end{center}
\end{table*}
\subsection{Cosmic muons induced dead time}
The cosmic-ray muon flux underwater can be deduced from data at
different depths and extrapolated as a function of equivalent
water depth where $10^5$~g/cm$^2$ = 1,000~mwe (meters of water equivalent). 
Using the depth--intensity relation for the total muon flux with a
flat overburden given in~\cite{meihime}, we derive the muon rate 
in a detector containing $10^{34}$ free protons having a section
exposed to muons of $1.48 \; 10^{8} \; {\rm cm}^2$ 
(cylinder of r=23~m radius and l=96.5~m length). 
At water depths of 500~m, 1,000~m, 2,000~m, and 4,000~m we obtain a 
muon rate of 560, 110, 8, 0.3~Hz, respectively. Vetoing the detector
for 200~$\mu$s after each muon would thus lead to respective 
muon-induced dead times of 11\%,  2.2\%,  0.15\%, and 0.006\%. From this
data we derive the minimum operating depth our module
         at around 0.5 km, not yet accounting for background.
We include this depth-dependent dead time in our sensitivity estimates.
Reducing the detector's dead time at shallower depth is possible by
subdividing the module in optically decoupled compartments. 
\begin{figure*}
\begin{center}
	\includegraphics[scale=0.45]{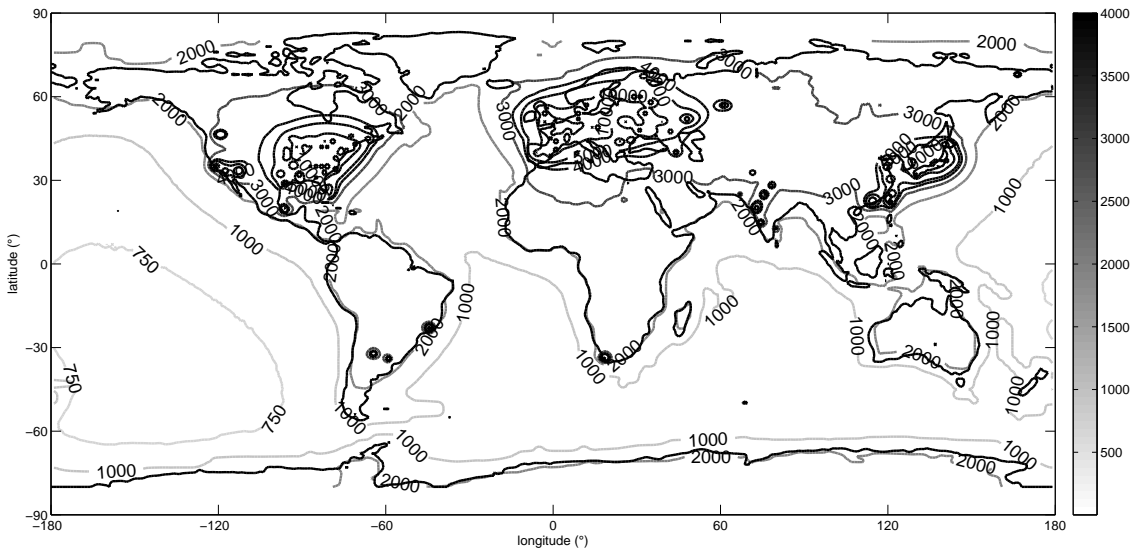}
	\caption{\label{bkg_noT_2500m} Maps illustrating the neutrino
          and non-neutrino background events that would be detected in
          a $10^{34}$ free protons detector (1 MeV energy threshold),
          operating for half a year at a depth of 2,500~m. At distances
          greater than 1,000~kilometers from nuclear power station
          clusters, the rate is dominated by geoneutrino events (see
          Section~\ref{geonu}). This background could be rejected by 
          setting an analysis threshold $E_{\rm vis} > 2.6$~MeV.}
\end{center}
\end{figure*}
\subsection{Correlated backgrounds}
Cosmic ray muons will be the dominating trigger rate at the depth of
our detectors. Their very high energy deposition corresponds to about
2~MeV per centimeter path length and provides thus a strong
discrimination tool. They induce the main source of dangerous events,
cosmogenic activity and fast neutrons, which mimic the
neutrino signal. 
\subsubsection{Muon induced cosmogenic activity}
\label{bkgcosmo}
Energetic muons can interact with carbon nuclei and produce by
spallation radioactive isotopes such as $^8$He, $^9$Li and $^{11}$Li.
These nuclei are unstable and decay emitting an electron and a neutron,
thus perfectly mimicking the signal from an antineutrino interaction ;
moreover the long lifetimes of these nuclei, a few 100~ms, complicate
the task of identifying them since they occur long after the muons that
created them.  
This background is considered to be the most
serious difficulty to overcome in the SNIF concept, and drives the operating
depth of the detector.  When this background is reduced - roughly
below 3,500~m  of water - $^8$He, $^9$Li and $^{11}$Li decays could be
identified through a four-fold coincidence ($\mu\to n\to\beta\to n$)
characteristic signature.

In order to estimate the cosmogenic backgrounds we used the rates 
experimentally measured by the Kamland~\cite{kamlandCosmo2010} 
 and Borexino~\cite{borexinogeonu} collaborations. 
The main detector features as well as muon induced backgrounds are provided 
in Table~\ref{tabsnifsbkg}. 
The production rate of cosmogenic radioisotopes is proportional to the
muon flux ($\Phi_\mu$), the cross section 
$ \sigma_{tot}(E_\mu) \sim E^{0.73}_\mu$~\cite{hagner2000}, and the
total number of carbon nuclei. We start from the muon flux predictions 
at the flat equivalent overburden of the KamLAND (2050 m of water
equivalent) and Borexino (3,050~m of water equivalent) locations. 
We then rescale the backgrounds to
 $10^{34}$~H.year for different depths according to the total
 muon flux, and energy formulae given in~\cite{meihime}. Estimates are 
corrected according to the different carbon composition in Borexino, 
KamLAND and SNIF liquid scintillators. Results for a $10^{34}$
H.year target deployed at a depth of 2,500~m are presented in 
Table~\ref{tabsnifsbkg}. An additional selection of events with
$E_{\rm vis}>2.6$ MeV would reject 23\% of the cosmogenic
background, according to the simple spectrum shape presented in~\cite{mention}.
In this work we envisage a further possibility of eradicating the cosmogenics
backgrounds by vetoing the detector after each muon for a long enough time.
We veto a 3-m-radius cylinder around each muon track for 600~ms (3~times the
$^8$He, $^9$Li decay time periods). 
Neglecting the veto inefficiency we find that this technique could 
only be effective at operating depth greater than 1,500~m to preserve
 a dead time below 10\% in a 138,000 ton detector. We thus neglect the 
cosmogenics backgrounds from this depth on.  Below 1,500~m we 
considered the KamLAND rescaling from~\cite{kamlandCosmo2010}.
The evolution of the predicted cosmogenics background in SNIF with 
respect to the detector depth is presented on Figure~\ref{snifbkg}.  

For completeness we note that a water Cherenkov detector is less 
affected by this specific background~\cite{sksolar} because of the
lower yield of these nuclei when spallation reactions occur with oxygen nuclei.
\subsubsection{Muon induced fast neutron activity}
\label{fastn}
An important source of background comes from neutrons produced in the 
surrounding of the detector by cosmic ray muon induced hadronic
cascades. The difficulty is that the primary cosmic ray muon may not 
penetrate the detector, being thus invisible. This is especially true
for a small detector. These processes produce arbitrarily high energy
 neutrons inducing proton recoils mimicking the prompt signal, and
 then producing the coincident neutron signal after thermalization and
 capture. Such a sequence can mimic a $\nuebar$ event. 
At several hundred meters of water equivalent, 
muon-induced neutron production can be
fairly well estimated from the results of previous underground
experiments, like KamLAND~\cite{kamlandOsc2008}  and 
Borexino~\cite{borexinogeonu}. Fast neutrons can indeed be produced
 by muons either crossing the inner stainless steel vessel or interacting in the
water around the detector. We estimate the rate of fast neutrons by
scaling the KamLAND and Borexino results according to the procedure 
described in Section~\ref{bkgcosmo}. Results for a $10^{34}$
proton.year target deployed at a depth of 2,500~m are presented in 
Table~\ref{tabsnifsbkg}. An additonal selection of events with
$E_{\rm vis}>2.6$ MeV would reject 31\% of the fast neutrons
background, assuming a flat energy spectrum.

For SNIF we only consider the case of Borexino and we reduce 
 the neutron production from the rock (4/5~of the total rate) by a
 factor of 2.7~in order to correct for the lower density of water.
Contrary to smaller detectors like KamLAND and Borexino, in very large
detectors like SNIF the fast neutron background could be considered as
a surface background. We thus assume that the fast neutrons induced by
muons crossing the bulk of the detector could be perfectly tagged in a
very large liquid scintillator detector (neglecting tiny veto
inefficiencies). The dominant fast neutron component induced by the
water surrounding the detector can only be detected at distances less
than 3 meters from the detector inner vessel walls. Further inside the
detector fast neutrons will have been significantly slowed.  This
lowers the estimated background by 70\%. The evolution of our
predicted fast neutron rate with respect to the detector depth is
depicted on Figure~\ref{snifbkg}.  We assume an uncertainty of 10\%,
achievable within the next 30~years.
\section{Detecting a clandestine activity}
\label{nuclearreactordetection}
In this section we detail the statistical method used to
decide whether a signal of undeclared activity is seen above the
known sources and backgrounds, at a given confidence level.
\subsection{Defining a decision threshold}
\label{lc}
Let $b$ be the background, i.e. the average number of events occurring in the
detector in the absence of any clandestine activity. Either from past
measurements, or theoretical calculations or other
input, $b$ is known with a certain uncertainty $\sigma_b$. 
We will treat this error as Gaussian. Now let $o$ be the
number of events actually observed in the detector.  
The first question one can ask
is whether $o$ is compatible with a fluctuation of the background, or
whether it is too high and could therefore be a sign of suspicious activity.
This is a well known problem, studied in great details in
e.g.~\cite{currie}. We will simply reproduce some of the explanations
and calculations therein. Following Currie's notation, we will write $L_C$ the
\textit{decision threshold} or \textit{critical level}: it is the
observed number of events above which the operator will declare having
observed a positive signal, i.e. detected a possible clandestine reactor.
Of course this level depends on the rate
of false alarms (incorrect reported detection while no clandestine
reactor is present, also known as type~I error) that one is willing to tolerate a priori. 

In the absence of any clandestine reactor, $o$ follows a Gaussian
distribution, with mean $b$ and error $\sqrt{b+\sigma_b^2}$. Given a
certain confidence level $\alpha$, 
$$o<b+k_\alpha\sqrt{b+\sigma_b^2},$$ where $k_\alpha$ is the
$\alpha$ quantile of the normal distribution.
Consequently, 
\begin{equation}
L_C=k_\alpha\sqrt{b+\sigma_b^2}.
\end{equation}
 If, having observed
$o$ events, $o-b>L_C$, then the detector will report the presence of a
clandestine reactor, with a pre-determined false alarm rate $1-\alpha$.

This criterion has the advantage of being directly applicable to data, and is used to make
a decision on whether to take further action. It depends solely on the mean and uncertainty 
on the background, and the chosen confidence level.
Note that at this stage there is no localization of the detector. The purpose of $L_C$
is to decide whether the observation signals suspicious activity or not.
With the definition above, $L_C$ is a number of events. 
If the distance from the reactor is known, $L_C$ can
be converted to a power (see section~\ref{sec:nurates}): 
this power is the maximum power that
would yield a signal consistent with the background.
We will often convert $L_C$ to either power or distance in
what follows. 
Conversely, if the power of the clandestine reactor is known, we can use the
considerations of section~\ref{sec:nurates} to convert $L_C$ to a
maximum distance below which the operator would declare having observed
a positive signal.

\subsection{Defining a detection limit}
\label{ld}
Following~\cite{currie}, we can also determine $L_D$, the
\textit{minimum detectable signal}, viz.~the minimal amount of signal
that could a priori be detected.  For this purpose it is first
necessary to determine $L_C$ at a certain false alarm rate $\alpha$,
as explained in the previous section. We then define a second
confidence level $\beta$, controlling the amount of type~II error that
we tolerate: $\beta$ is the probability that an existing reactor would
be missed by our method.
With these definitions, $L_D$ is the average amount of signal that would lead to
detection (i.e.~observation of more than $L_C$ counts) with probability $1-\beta$.  
The solution to this problem is found in \cite{currie}:
\begin{equation}
L_D =  L_C + \frac{k_\beta^2}{2}\left(1+\sqrt{1+\frac{4L_C}{k_\beta^2}+4\frac{L_C^2}{k_\alpha^2 k_\beta^2}}\right).
\end{equation}
With these two quantities, we can explore the sensitivity of our detection method.
As for $L_C$, $L_D$ can either be converted to a power (if the distance from
the reactor is known) or to a distance (if the power of the reactor is known).
At known distance, the resulting power is the minimum power detectable a priori
with type~I and~II error rates of $\alpha$ and $\beta$ respectively.
At known power, the resulting distance is the maximum distance at which
this power would a priori be detected, with type~I and~II error rates of $\alpha$ 
and $\beta$ respectively.

\subsection{Impact of known nuclear reactors }
In this section we study $L_C$ and $L_D$ (or their conversions to
power/distance), as a function of the
detector mass, exposure time, clandestine reactor thermal
power, and the known nuclear reactor neutrino rate at any 
Earth location.

Let us introduce a new luminosity unit, called the r.n.u. (for reactor
neutrino unit) and defined as
$1~\text{r.n.u.}=0.197\;10^{30}~\text{MeV}$. With this unit, an
experiment taking data for $T$~years with a total clandestine nuclear
power of $P$~GW$_{\rm th}$. and with $N\;10^{34}$~free protons inside the
target has a luminosity ${\cal L}=T\,P\,N~\text{r.n.u.}$ The expected
number of events, $N(L)$, at a distance $L$ from a reactor, assuming
no\,-\,oscillation, is
\begin{equation}
\label{eq:rnu}
\begin{split} 
N(L)= & \frac{\left<\sigma_f\right>}{4\pi\left<E_f\right>}\frac{\cal
   L}{L^2} \simeq 230 \;  \left(\frac{T}{0.5\text{[y]}}\right)  \\
&  \left(\frac{P}{100\text{[MW]}}\right)  \left(\frac{N}{10^{34}}
\right)
 \left(\frac{1}{L\text{[km]}^{2}} \right)
\end{split}
\end{equation}
%
The rate in equation~\ref{eq:rnu} is then corrected for neutrino oscillations in all our calculations.
\begin{figure*}
\begin{center}
	\includegraphics[scale=0.5]{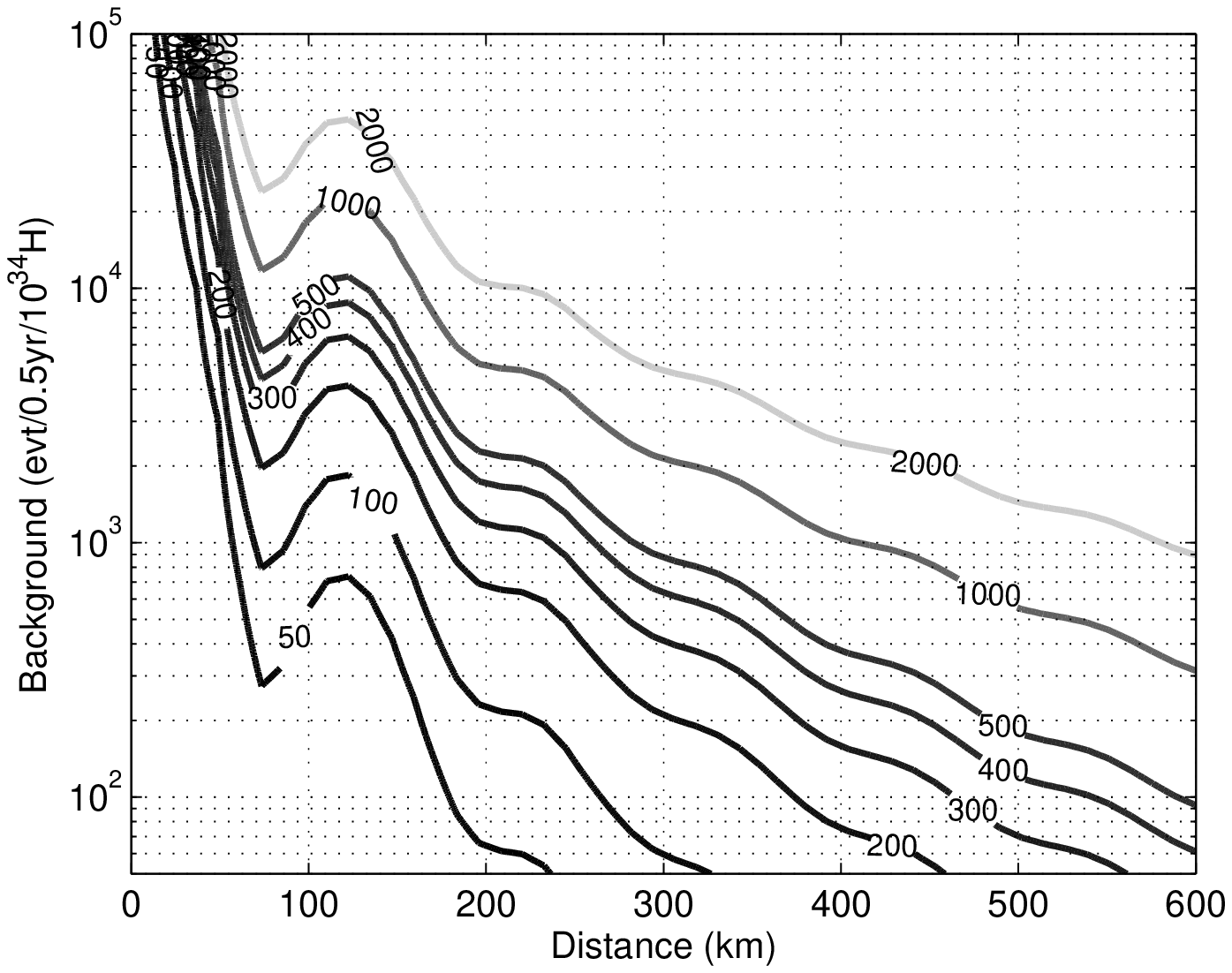}
	\includegraphics[scale=0.5]{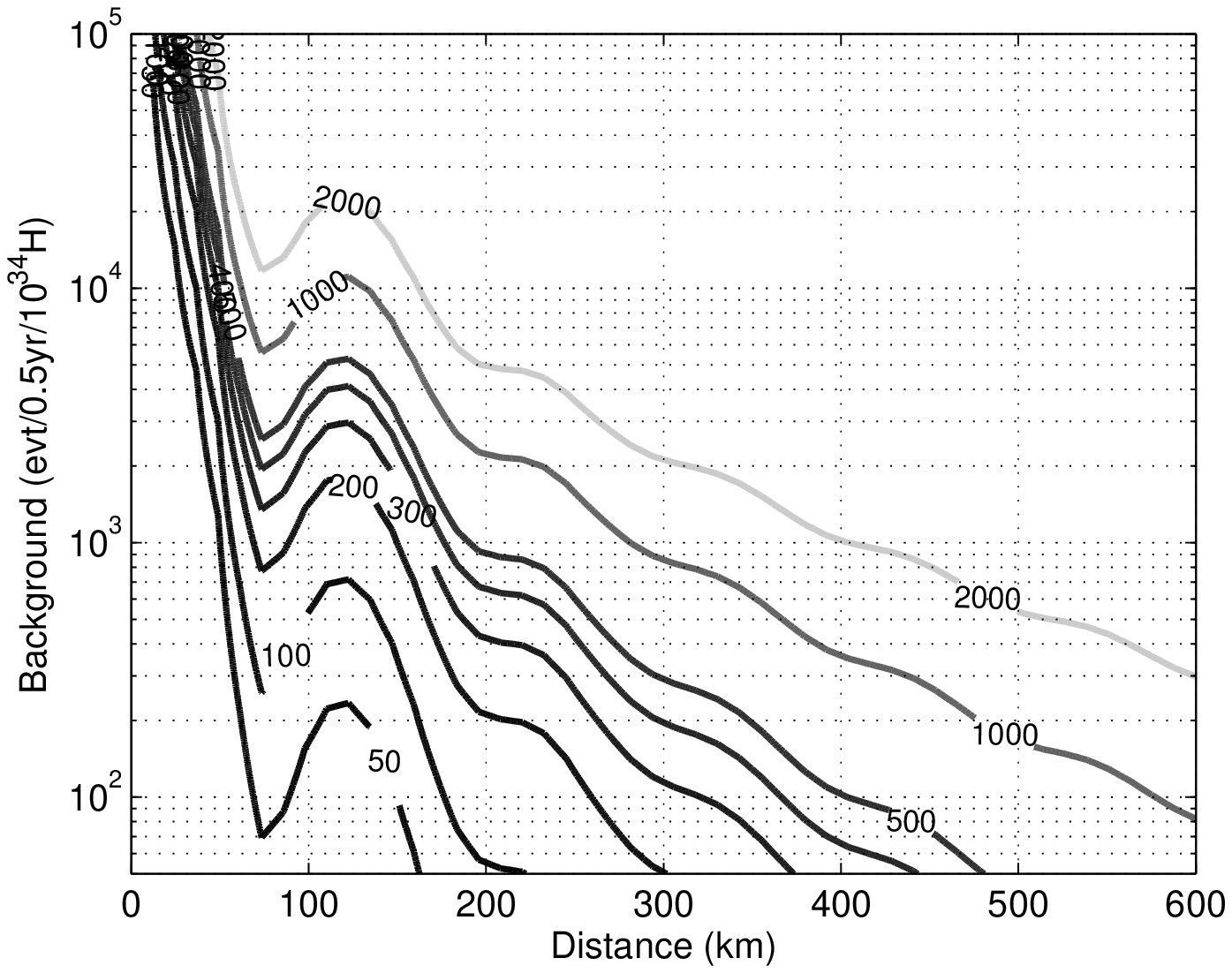}
	\caption{\label{abaques_nubkg} 
          Clandestine nuclear reactor power (MW) detectable with the
          neutrino method accounting only from known neutrino
          sources. (left) Maximum power (in MW) of
          an undeclared nuclear reactor consistent with the known neutrino
          sources as a function of the reactor distance (km) and
          neutrino rate from commercial nuclear power stations,
          as described in section \ref{lc}. We set a 10\% false alarm
          tolerance.  (right) Minimum power (in MW) of an undeclared
          nuclear reactors that could a priori be detectable by the
          neutrino method as a function of the neutrino background and
          the reactor distance (km), as described in section
          \ref{ld}. The false alarm rate is 10\% and the probability
          of missing an existing reactor is 10\%. We consider an
          exposure of  $0.5\;10^{34}$ H.y. Wiggles
          in the range 50-150 km are due to neutrino oscillations.}
\end{center}
\end{figure*}

Figure \ref{flux} shows the neutrino rate world map at an operating
depth of 4,000~m. Correlated backgrounds, as described in
Section \ref{nonnubkg}, turn out to be negligible at this depth.  We
apply a 2.6 MeV visible energy threshold, rejecting accidental and
geoneutrino backgrounds. 
Three representative cases of the commercial reactor neutrino rate levels
can be identified: a low rate area, the southern hemisphere, 
where the detector would detect $<500$ events in 6~months, 
a medium rate area corresponding to $500-1,000$ events, and a
 high rate area corresponding to $>1,000$ events or more, 
near clusters of nuclear power stations in Europe, Japan, and North America. 

Figure \ref{abaques_nubkg} (left) provides the maximum power of an undeclared
nuclear reactor consistent with the known neutrino sources. The
graphics reads as follows: the horizontal axis is the distance between
the hypothetical clandestine reactor and the neutrino detector, 
the vertical axis is the known neutrino rate from commercial
nuclear power stations that could be related to an Earth location. 
The lines represent the iso-power (MW) contours quantifying the
maximum power consistent with the known sources. In all what follows
we set a 10\% false alarm tolerance (type~I error) as described in Section \ref{lc}. 
In medium background conditions we note that a 300~MW reactor
could be inferred after 6 months of observation 
with a single 138,000~ton liquid scintillator detector located 300~km away.

Figure \ref{abaques_nubkg} (right) provides the minimum power of
 a nuclear reactor detectable by the neutrino method as a function 
of the neutrino background and the clandestine reactor distance. 
The probability of missing an existing reactor, or type~II error as
described in Section \ref{ld}, is set to 10\% for the rest of this article.
\subsection{Impact of non reactor neutrino backgrounds}
\label{bkgdiscuss}
In this section we discuss for the first time the impact of the
geoneutrinos and non-neutrino backgrounds on the sensitivity of 
the neutrino method to detect undeclared nuclear fission activities. 
\begin{figure*}
\begin{center}
	\includegraphics[scale=0.5]{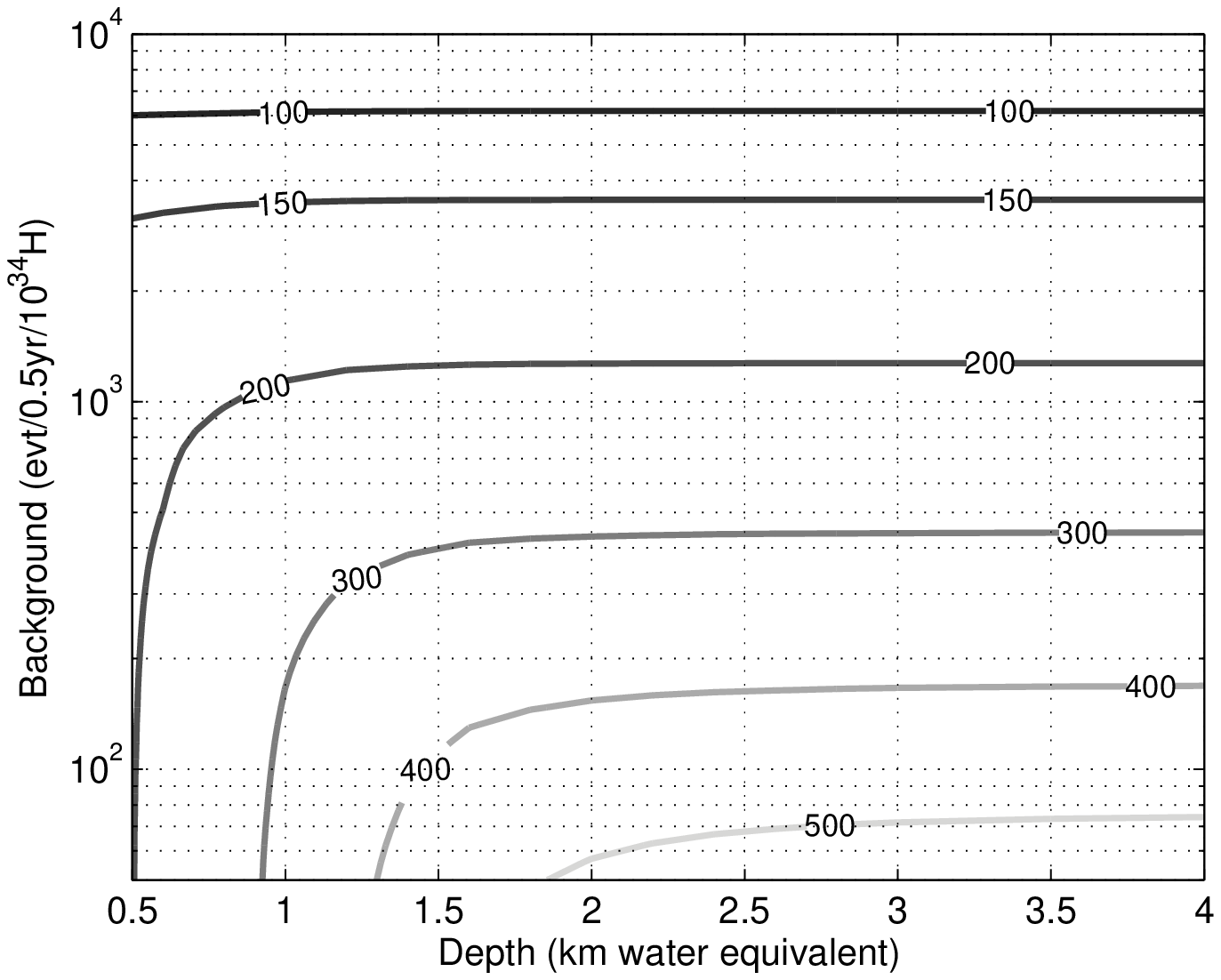}
	\includegraphics[scale=0.5]{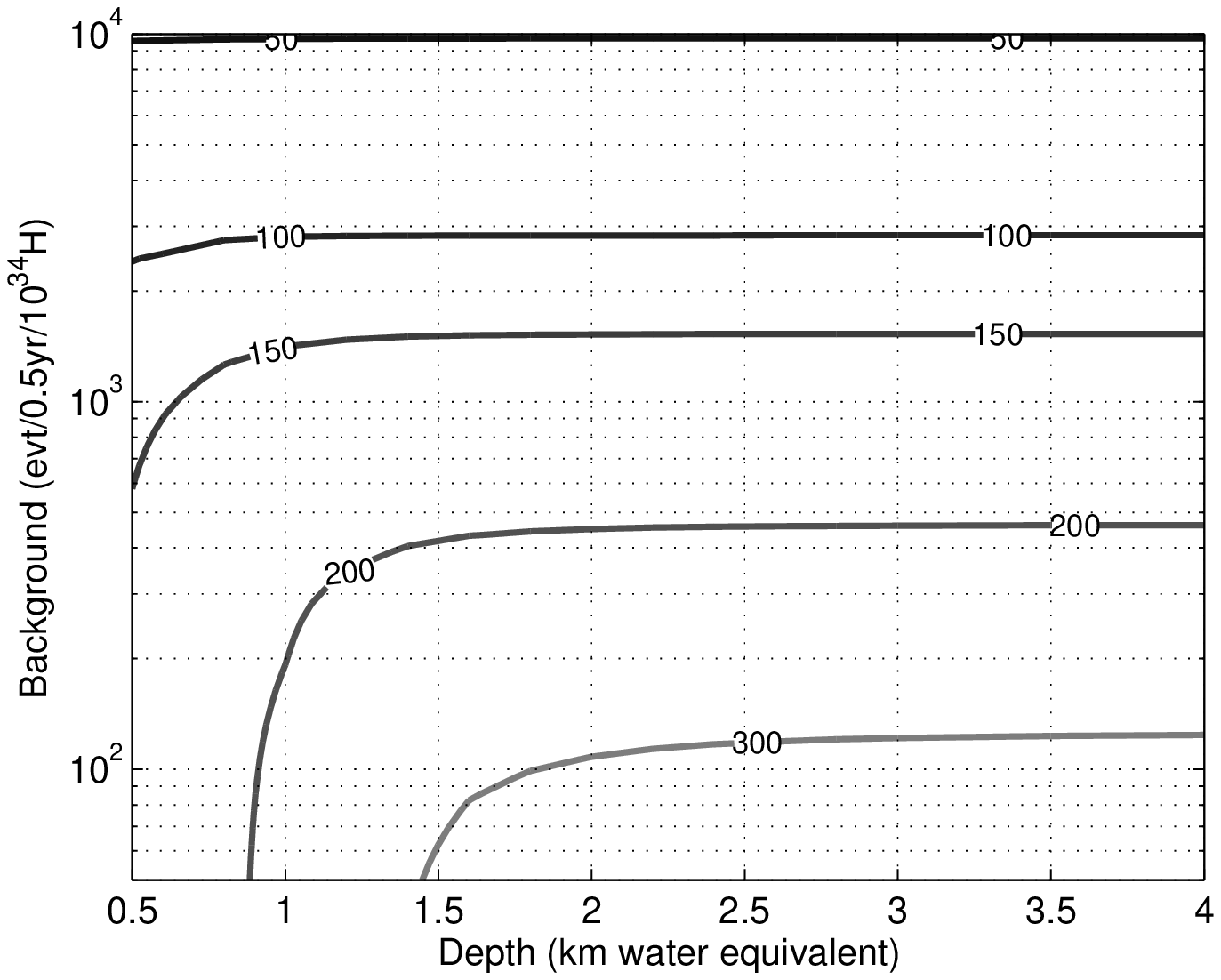}
	\caption{\label{abaques_depth} 
          Impact of the non-neutrino backgrounds on the sensitivity to
          undeclared nuclear fission activity. The horizontal axis
          provides the operating depth of the neutrino detector, in
          kilometers of water (kme). The vertical axis represent the
          known neutrino source rate, for an exposure of
          $0.5\;10^{34}$ H.y.
          Assuming a 300~MW undeclared reactor the panel contours 
          describe the iso-minimum distances (km) at which the signal 
          from the clandestine reactor is consistent with the
          expectation. At closer distances the neutrino detector could
          tag the clandestine activity. The right panel shows the
          maximum distance (km)  at which one could a priori verify
          the consistency of the signal with the 300~MW 
          reactor hypothesis.}
\end{center}
\end{figure*}

The map of Figure~\ref{bkg_noT_2500m} illustrates the background rates
 at a depth of 2,500~m that would arise by lowering the detection
 threshold to 1 MeV (visible energy). This can be compared with 
Figure \ref{flux}. At distances of more than a thousand kilometers 
from nuclear power station clusters the rate is dominated by
geoneutrino events. Extrapoling from Figure \ref{abaques_nubkg}
(left) we see that geo-neutrinos would prevent the detection of any 
reactor of $<$1~GW with a detector located at $>$ 300~km in the
$medium$ commerical neutrino rate regions.
This further justifies the choice of setting a 2.6~MeV
visible energy threshold.
 
The influence of backgrounds on the sensitivity to rogue activity is
illustrated on Figure~\ref{abaques_depth}, as a function of the
operating depth and of the expected neutrino rate from
nuclear power stations. In this case we assume the
existence of a 300~MW rogue reactor. 
We first notice that for known neutrino source rates of more than a few
thousand events the sensitivity is weakly affected by the operating
depth below 500~mwe since the non-neutrino background rate remains a 
small fraction of the total neutrino-like rate.

The evolution of the clandestine
activity detectability with respect to the operating depth is
illustrated on Figure \ref{abaques_depth} (left).  
In medium background conditions we note that a 300~MW reactor
could be inferred after 6 months of observation 
with a single 138,000~ton liquid scintillator detector located 300~km
away, only if the depth is greater than 1,500~m. 
We note that the detection distance would be degraded 
to 200 km at a depth of 600~mwe.
According to our background model we conclude it is not 
necessary to deploy a detector module below 2,000~m.
This is one of the main results of this article. 
\begin{figure*}
\begin{center}
	\includegraphics[scale=0.5]{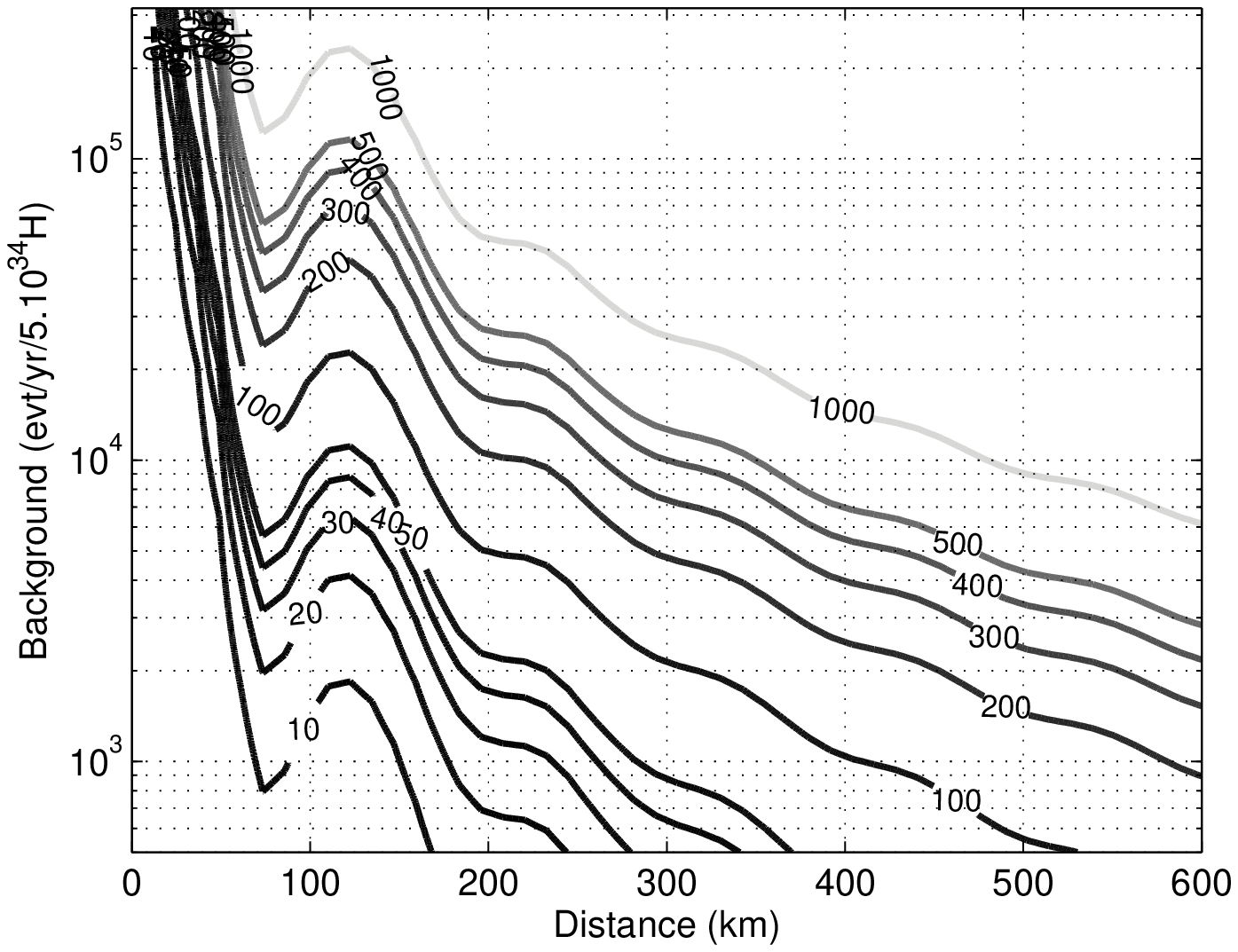}
	\includegraphics[scale=0.5]{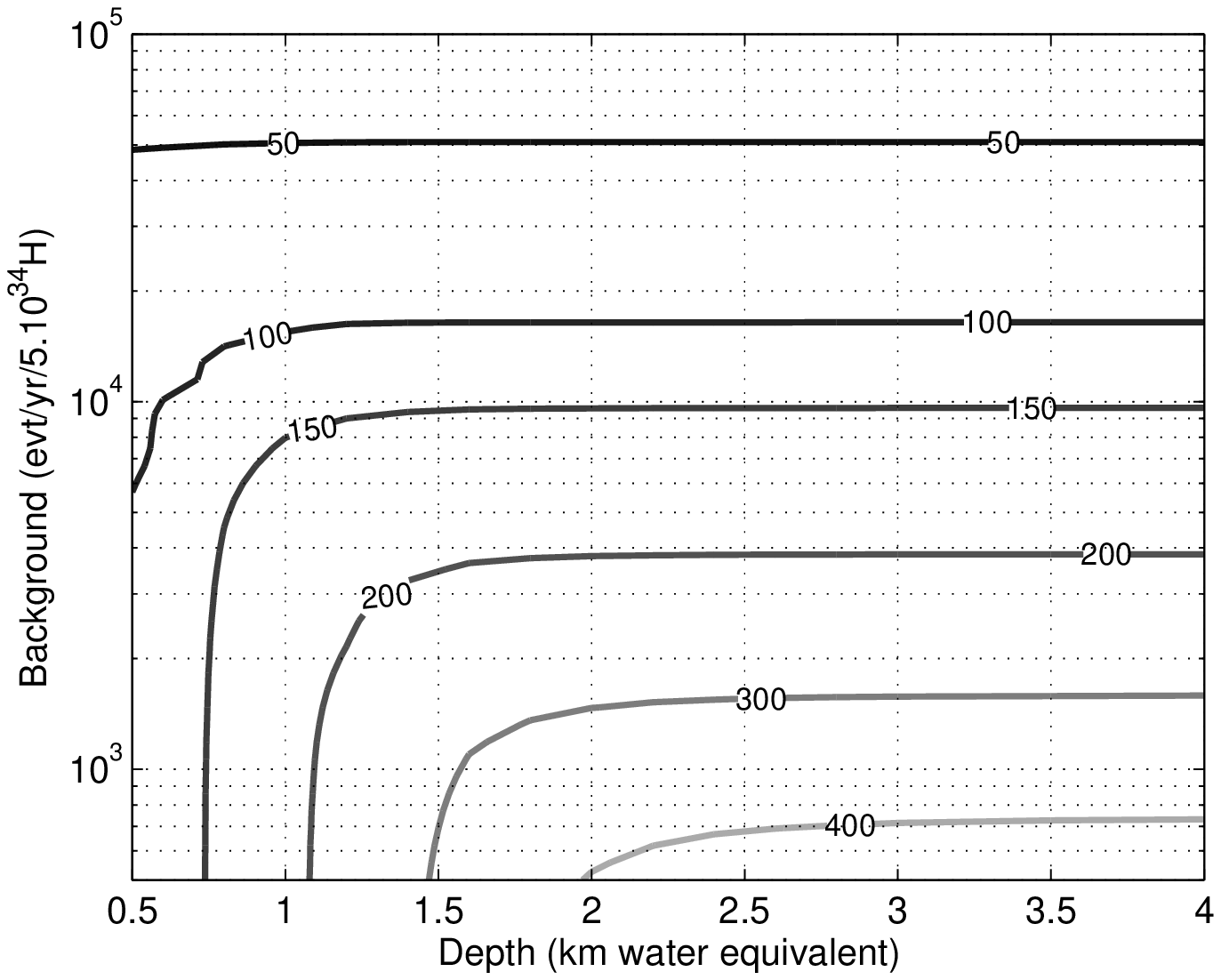}
	\caption{\label{abaques_megaexpo} 
           Remote surveillance of low power research reactors with 
           five $10^{34}$ free protons detector modules operating for 
           one year. The left panel shows the iso-maximum power
           lines (MW) consistent with expectation, as a
           function of the detector-reactor distance (km) and the
           known reactor neutrino rate. 
           The right panel describes the iso-minimum distance contours
          (km) at which a 75~MW rogue reactor is
          consistent with all background expectation as a function of
          the detector depth (km) and known reactor neutrino rate.
        }
\end{center}
\end{figure*}
\subsection{Scaling up the detection volume}
\label{scalingup}
Having developed a compehensive background model we now study 
the limitation of the neutrino method by scaling up the exposure. This
could be done either by enlarging the size of the detector modules, or
by operating several detector modules simultaneously, or by increasing
the operating period. The potential application could be the surveillance
of low power research reactors of a few tens of~MW.
Let us consider an exposure of $5\times 10^{34}$ H.year, ten times more than
our baseline case. This could be realized
with five detector modules of 138,000~tons operating for 1~year. 
Results not considering non-neutrino backgrounds are displayed on 
Figure \ref{abaques_megaexpo} (left). For a typical southern
hemisphere location (3,000~events from known neutrino sources) we conclude
that this system would be sensitive to a 50~MW reactor from a distance of 200~km.
The right panel of Figure \ref{abaques_megaexpo} (left) shows how 
the sensitivity evolves with the operating depth, still allowing the
detection of a 75~MW reactor at a distance of 150~km with detectors 
operating under 750~m.
\section{Clandestine reactor localization}
\label{reactorloc}
The strategy developed in this article is to deploy a detector as
close as possible to a suspicious area to find evidence of clandestine
activity.  If any evidence of clandestine activity is found,
additional detectors have to be deployed with the objective of finding
the clandestine reactor's location. Four detector modules (i=1,2,3,4)
operating at four distinct locations at longitude and latitude
($\lambda$,$\phi$)$_{i}$ with a positive decision threshold are
necessary to isolate a unique location, inferring in addition the
reactor's power. We present here a simple optimization algorithm
providing an approximate position of the presumed clandestine reactor
as well as confidence levels of the clandestine reactor location, for
the best fitted thermal power. For simplicity we assume the reactor to
be constantly running.

Let's assume the presence of an undeclared reactor with a power P (MW)
at a location ($\lambda$,$\phi$)$_R$. S$_i$ is the neutrino
signal induced by the rogue reactor in the detector $i$,
drawn according to a Poissonian distribution of mean S$_i$. B$_i$ is
the expected background, including known neutrino sources as well as
non-neutrino backgrounds. Background uncertainties are taken from Sections
\ref{neutrinobkg} and \ref{nonnubkg} and are added in quadrature.  O$_i$
is the observed value in the detector $i$, following a Gaussian of
variance O$_i$ (statistical error). The triplet
($\lambda$,$\phi$,P)$_R$ 
is estimated by minimizing the $\chi^2$ function:
\begin{equation}
\chi ^2 = \sum_{i}
\frac{(O_i-S_i(\lambda,\phi,P)_R-B_i)^2}{B_i+S_i(\lambda,\phi,P)_R} + \frac{(P-P_{exp})^2}{\sigma_{Pexp}^2} 
\end{equation}
$P_{exp}$ and $\sigma_{Pexp}$ are set to 300~MW and 50~MW respectively;
in practice they would be set to values that best represent typical
rogue reactors.
                                   
With four detectors the $\Delta\chi ^2$ function follows a $\chi ^2$
distribution with 3~degrees of freedom. However we fit the thermal power
at each point on the contour map, allowing to derive the
($\lambda$,$\phi$)$_{CL}$ confidence intervals at the CL=68.3\%
(1$\;\sigma$) and CL=95.4\% (2$\;\sigma$) by selecting respectively 
the $\Delta\chi^2= \chi^2(\lambda,\phi)-
\chi_{min} ^2$=2.30 and 6.18 areas (2~degrees of freedom since $P_R$
is fitted at every location).
The second term on the right-hand side disfavors large powers
inconsistent with low power reactors. It eliminates degenerate solutions
located too far away from the region of interest. Any further
information providing the thermal power would greatly enhance the
localization algorithm, with a possible reduction of the number of
detectors.

We notice here that a measurement of the neutrino energy spectrum
distorsion due to neutrino oscillations in the undeclared reactor's spectrum
observed with a detector located 70--150~km away could improve the precision of the
localization. This effect, already measured by the KamLAND
detector~\cite{kamlandOsc2008}, would however require much higher statistics from
the undeclared reactor. We will thus neglect it in our study.

The accuracy and robustness of the method depends on the
geographical location of the rogue activity. We now consider
three distinct cases: a penisula, an island, and a flat shore. 
Though the true latitude and longitude coordinates of our examples 
are hidden they correspond to real cases.
In each case the reactor is placed at the origin of each of our coordinate system.
In each case the known neutrino sources amount of several hundreds 
of counts per $10^{34}$ H.y corresponding to intertropical zones.
\subsection{Peninsula}
\begin{figure}[!h]
\begin{center}
	\includegraphics[scale=0.4]{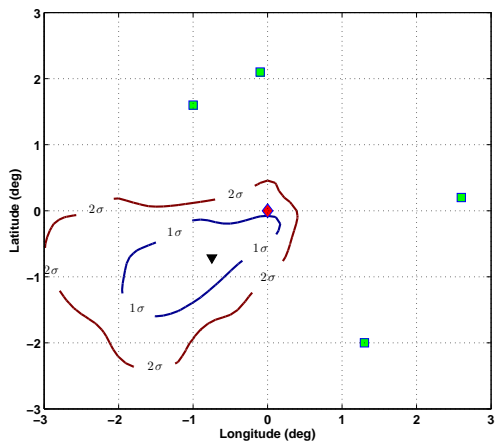}
	\caption{\label{peninsula_locate} Localization of a
          clandestine reactor located on a peninsula. Four detectors (squares)
          surround a 300~MW$_{\rm th}$ reactor (diamond). 
          The triangle shows the reconstructed position at
          the best fitted power. The 68.3\% (1 $\;\sigma$) and  95.4\%
          (2$\;\sigma$) confidence levels are displayed.
          The thermal power at best fit is $P=429$ MW$_{\rm th}$. 
          }
\end{center}
\end{figure}

\begin{figure}[!h]
\begin{center}
	\includegraphics[scale=0.45]{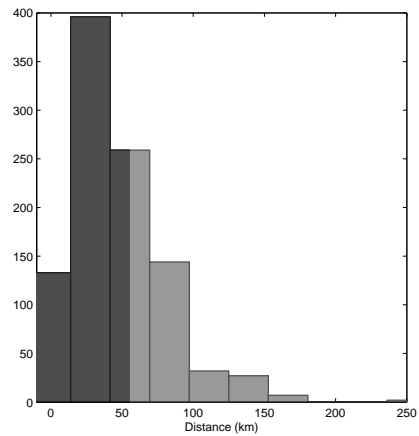}
	\caption{\label{peninsula_recoD} 
          Distribution of the distance (km) between the true and the
          reconstructed positions for 1,000~trials of the peninsula
          experiment. In this particular case the detectors operate
          at a depth of 1,500~m. 
          The position is reconstructed within $D_{68\%}$=55~km 
          (dark gray area) in 68\% of the cases. }
\end{center}
\end{figure}

The first case assumes a
300~MW$_{\rm th}$ clandestine reactor located on a peninsula. 
Four detectors are deployed 1,000~m underwater~209 to 264~km away 
from the clandestine reactor. They operate for 1 year. Known reactor 
neutrino rates provide 631~events in each detector on average, 
to be compared with a mean rogue signal of 76~events.
The clandestine reactor is clearly 
detected by three of the four modules (see Table~\ref{tab_loc_peninsula}). 
\begin{table}[htbp]
\fontsize{8}{10}\selectfont
\begin{center}
\scalebox{1}{
\begin{tabular}{c|c|ccc|cc}
\hline
\textbf{($\lambda$,$\phi$)} & \textbf{Distance} & \textbf{$<$S$>$} & \textbf{$<$B$>$} & \textbf{O} & $L_c$ &  \textbf{O}-\textbf{$<$B$>$}    \\
\hline
\textbf{(-1.0$^\circ$,+1.6$^\circ$)} & 209 km& 101& 643& 749&  54  & 106\\
\textbf{(+0.1$^\circ$,+2.1$^\circ$)} & 234 km& 86 & 645& 714& 54  & 69\\
\textbf{(+2.6$^\circ$,+0.2$^\circ$)} & 286 km& 52 & 628& 627&  53 & -1\\
\textbf{(+1.3$^\circ$,-2.0$^\circ$)} & 264 km& 63 & 610& 714&  53 & 104\\
\hline
\end{tabular}
}
\caption{Four detectors (each containing  10$^{34}$ free protons) 
  are operating undersea at a depth of 1,000~m, for 1~year,
  about $250$ km away from a 300~MW clandestine reactor
  located on a peninsula. 
  S is the rogue signal. B is the sum of the counts from known nuclear
  reactors and non-neutrino backgrounds. 
  O is the observed value according to a random
  experimental trial. $L_c$ is the decision threshold
  value. \mbox{$O-L_c>B$} is the detection criteria.}
\label{tab_loc_peninsula}
\end{center}
\end{table}
Figure \ref{peninsula_locate} demonstrates the ability to detect 
and locate the 300~MW$_{\rm th}$ clandestine reactor with four detector
modules located at an average distance of 250~km.
In order to assess the perfomances of the neutrino method we draw 1,000~random 
trials of the peninsula experiment at various detector operating depths: 500, 750, 1,000,
1,500, 2,500 and 3,500~mwe. We then reconstruct the reactor's position
and power.
Figure~\ref{peninsula_recoD} illustrate the reconstruction of the
reactor position with a detector immersed under 1,500~mwe. The spatial
resolution, estimated as the 68\% quantile of the distance between the
best fit position and the true position, is $D_{68\%}$=55~km. Similarily the mean
reconstructed thermal power is 241$\pm$44~MW. Results are various depth
are presented on Table \ref{tab_trials}.
\subsection{Island}
\begin{figure}[!h]
\begin{center}
	\includegraphics[scale=0.4]{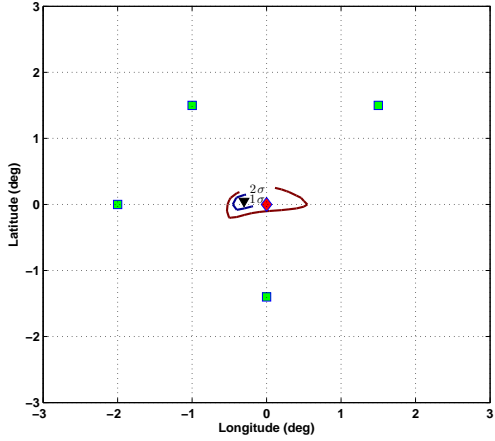}
	\caption{\label{island_locate} Localization of a
          clandestine reactor located on an island. Legends are
          similar to those of Figure \ref{peninsula_locate}. 
          The thermal power at best fit is $P=288$~MW$_{\rm th}$.  
          }
\end{center}
\end{figure}

\begin{figure}[!h]
\begin{center}
	\includegraphics[scale=0.45]{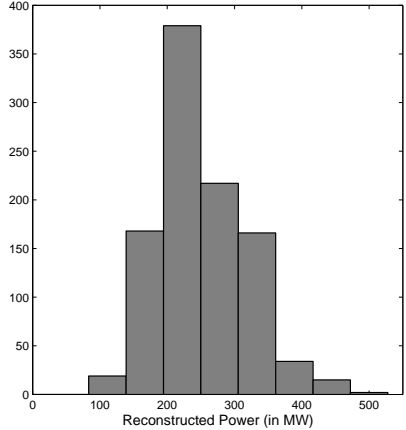}
	\caption{\label{island_recoP} 
          Distribution of the reconstructed power for 1,000~trials of
          the island experiment (1,000~mwe). 
         The mean reconstructed power is 254$\pm$63~MW.}
\end{center}
\end{figure}

The second case assumes a
300~MW$_{\rm th}$ clandestine reactor located on an island.  
Four detectors are deployed at a depth of 1,000~m, between 156~and 235~km from the
clandestine reactor. They operate for 1~year.  All the known
neutrino sources provide 750~events on average, to be compared
with a mean rogue signal of 127~events. The clandestine reactor is
unambiguously detected by the four modules (see Table
\ref{tab_loc_island}). 
In the example displayed on Figure~\ref{island_locate} its position is
well reconstructed, a few tens of kilometers from the true location.
The thermal power is well estimated, at 288~MW, attesting to the
potential performance of the neutrino method in a favorable case where
detectors surround the suspicious location.
\begin{table}[htbp]
\fontsize{8}{10}\selectfont
\begin{center}
\scalebox{1}{
\begin{tabular}{c|c|ccc|cc}
\hline
\textbf{($\lambda$,$\phi$)} & \textbf{Distance} & \textbf{$<$S$>$} & \textbf{$<$B$>$} & \textbf{O} & $L_c$ &  \textbf{O}-\textbf{$<$B$>$}    \\
\hline
\textbf{(+1.5$^\circ$,+1.5$^\circ$)} & 235 km& 86& 772& 816&  57  & 44\\
\textbf{(0$^\circ$,-1.4$^\circ$)} & 156 km& 221 & 707  & 886& 55  & 179\\
\textbf{(-2.0$^\circ$,0$^\circ$)} & 221 km& 94 & 743 &  826&  56 & 83\\
\textbf{(-1.0$^\circ$,+1.5$^\circ$)} & 200 km& 107 & 779 & 922&  57 & 143\\
\hline
\end{tabular}
}
\caption{Four 10$^{34}$ free protons detectors are deployed at an
  average distance of 203~km from an undeclared reactor located on an 
  island. Each detector operates under water at a depth of 1,000~m, for 1~year. 
}
\label{tab_loc_island}
\end{center}
\end{table}
As for the peninsula case we randomly draw 1,000~trials of the island
experiment described above at various depths.
Figure \ref{island_recoP} illustrates the reconstruction of the
reactor's power with a detector immersed under 1,000~m of water.
The mean reconstructed thermal power is 254$\pm$63~MW. The spatial
resolution is $D_{68\%}$=43~km. Results at various depths are 
presented on Table \ref{tab_trials}.
\begin{table*}[htbp]
\fontsize{8}{10}\selectfont
\begin{center}
\scalebox{1}{
\begin{tabular}{c|cc|cc|cc}
\hline
 &  \multicolumn{2}{c}{\textbf{Peninsula}} & \multicolumn{2}{c}{\textbf{Island}}& \multicolumn{2}{c}{\textbf{Flat shore}}  \\
\textbf{Depth} & P/$\sigma$(MW) & $D_{68\%}$(km) & P/$\sigma$(MW) & $D_{68\%}$(km) &P/$\sigma$(MW) & $D_{68\%}$(km)\\
\hline
\textbf{500 mwe} &  211$\pm$188 & 271 & 249$\pm$153 & 196 &193$\pm$122 & 197 \\
\textbf{750 mwe} & 226$\pm$136 & 239 & 265$\pm$99 & 82 & 220$\pm$106 &186\\
\textbf{1000 mwe} & 234$\pm$86 & 128 & 254$\pm$63 & 44 & 236$\pm$83 & 182\\
\textbf{1500 mwe} & 241$\pm$44 & 55 & 244$\pm$31 & 32 & 266$\pm$61 & 113\\
\textbf{2500 mwe} & 243$\pm$38 & 49 & 243$\pm$29 & 36 & 266$\pm$58 & 16\\
\textbf{3500 mwe} & 242$\pm$30 & 47 & 242$\pm$29 & 32 & 264$\pm$56 & 16\\
\hline
\end{tabular}
}
\caption{Results of simulation of 1,000~trials of the peninsula,
  island, and flat shore virtual experiments for detector depths
  varying from 500~to 3,500~mwe. The determination of the power and 
 reactor location becomes more accurate as the depth increases until 
 backgrounds become negligible compared to known neutrino sources.}
\label{tab_trials}
\end{center}
\end{table*}
\subsection{Flat Shore}
\begin{figure}[!h]
\begin{center}
	\includegraphics[scale=0.4]{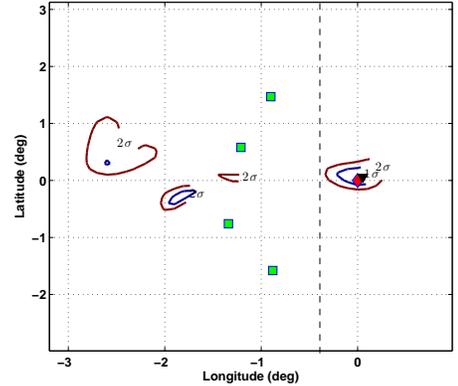}
	\caption{\label{flatshore_locate} 
          Localization of a
          clandestine reactor located on an island. Legends are
          similar to those of Figure \ref{peninsula_locate}. 
          The solutions to the west of the detectors are naturally excluded
          since they are located in the sea. 
          The thermal power at best fit is $P=250$~MW$_{\rm th}$.  
        }
\end{center}
\end{figure}

In the third case we consider a flat shore geometry spanning along the
latitude axis, at a longitude of about -0.4$^\circ$.  
We arbitrarily placed a 300~MW clandestine reactor 
a few tens of kilometers inland.
Four detectors are deployed at a depth of 1,000~m for 1~year. 
All the known neutrino sources provide 670~events on average, 
to be compared with a mean rogue signal of 165~events. 
The clandestine reactor is unambiguously detected by the 
four modules (see Table \ref{tab_loc_flatshore}). 
\begin{table}[htbp]
\fontsize{8}{10}\selectfont
\begin{center}
\scalebox{1}{
\begin{tabular}{c|c|ccc|cc}
\hline
\textbf{($\lambda$,$\phi$)} & \textbf{Distance} & \textbf{$<$S$>$} & \textbf{$<$B$>$} & \textbf{O} & $L_c$ &  \textbf{O}-\textbf{$<$B$>$}    \\
\hline
\textbf{(-1.2$^\circ$,+0.6$^\circ$)} & 146 km& 269& 676 & 910 &  55  & 233\\
\textbf{(-1.3$^\circ$,-0.8$^\circ$)} & 169 km& 168 & 661  & 766 & 54  & 105\\
\textbf{(-0.9$^\circ$,-1.6$^\circ$)} & 200 km& 107 & 655 &  746 &  54 & 91\\
\textbf{(-0.9$^\circ$,+1.5$^\circ$)} & 191 km& 117 & 687 & 766 &  55 & 79\\
\hline
\end{tabular}
}
\caption{Four detectors containing 10$^{34}$ free protons are located at an
  average distance of 177~km from an undeclared reactor located close
  to a flat shore. Each detector operates under water at a depth of 1,000~m, for 1~year. }
\label{tab_loc_flatshore}
\end{center}
\end{table}
In figure ~\ref{flatshore_locate},  we see 4~degenerate solutions for the
reactor location, the best fit position being reconstructed a few
kilometers from the true reactor location. We interpret it as the
impossibility to surround the true reactor location in this flat shore
configuration. Fortunately in this case the three solutions reconstructed to 
the west of the detectors lie in the sea and can thus safely
be rejected. 
A possible way of improving the localization is the reduction of 
the sources of backgrounds. At fixed reactor location one can only 
consider operating the detector deeper. Table \ref{tab_loc_flatshore}
shows how the localization accuracy varies with the detector depth. 
At a depth of 2,500~m or deeper the position of the reactor is
reconstructed within 15~km in 68\% of the cases. 
We notice that the reconstructed thermal power varies from 193$\pm$122~MW
to 264$\pm$56~MW when the operating depth ranges from 500 to 3500~mwe,
thus leading to a significant reduction of the uncertainty.

If more than one plausible solutions still exist with the four deep
detector deployement, additional detector modules, or moving the
existing modules in the fleet would be necessary 
to eliminate the degeneracies. 
\section{Conclusion}
\label{conclusion}
In the revival of the nuclear era new technologies may be used
to enforce the surveillance of nuclear activities. In this article we
discussed the futuristic option of using very large neutrino detectors
to detect clandestine nuclear reactors. 
In comparison with the previous studies of ~\cite{learned, guillian}  we 
considered detector modules of 138,000 tons, fitting inside an oil supertanker,
and using liquid scintillator technology
This corresponds to three times the volume of
the largest neutrino detector ever built in the 1990s~\cite{SK98}. 
The development of such a detector is not unrealistic 
within the next 30~years -- not taking into account financial constraints. 
The main technical challenge would be the deployment and 
operation of such a huge detector underwater. 

Our simulation concept, called SNIF, allows us to assess the
detectability of any clandestine nuclear reactor at any Earth
location. 
All known reactor neutrino sources have been
included in our simulation, including geoneutrinos. For the first time 
we provide a phenomenological model of non-neutrino backgrounds 
based on the scaling of recent reactor neutrino experiment results. 
In addition we modeled the non-neutrino background evolution as 
a function of the detector's operating depth. 
Beyond previous studies which only consider immersing detectors below 4000~m of water
~\cite{learned, guillian}, we found that large neutrino detectors could
also be deployed at depths ranging from 500~m to 2,000~m of waters.
As an example a 300~MW reactor could be detected
after 6~months of observation with a single detector located 
300~km away, operating at a depth greater than 1,500~m.
Using five detector modules for 1~year a 50~MW reactor could be
detected at 200~km.

Beyond detectability, we addressed the possibility of localizing
clandestine nuclear reactors with four detectors. The precision at
which we reconstruct the longitude, latitude, and
power of the reactor depends on the geographical situation. We 
considred three typical cases of reactors located on a peninsula, an
island, or on a flat shore. Localization of 300~MW nuclear reactors
within a few tens of kilometers is possible in such
conditions. In these cases the thermal power could be reconstructed
within 50~MW. However correlations between reconstructed power 
and location may lead to degenerate solutions that are can only be lifted with
additional detectors or extra information.

Our study attests that 138,000~ton neutrino detectors have the
capability to detect and even localize clandestine reactors from
across borders. However we conclude that clandestine reactor
neutrino detection would face formidable obstacles to implementation. 
\begin{acknowledgments}
We would like to thank F. Mantovani for providing us the geoneutrino fluxes map from the Earth model.
We are grateful to Jean-Luc Sida for sharing his insight and expertise.
\end{acknowledgments}

\bibliography{snif}

\end{document}